\documentclass[10pt,conference]{IEEEtran}

\usepackage{amsmath}
\usepackage{tcolorbox}
\usepackage{relsize}
\usepackage{hyperref}
\usepackage{balance}
\usepackage{import}
\usepackage{cleveref}

\usepackage{graphicx}
\usepackage{caption}
\usepackage{fontawesome5}
\graphicspath{{./images/}}

\usepackage{enumitem} 

\usepackage{tikz}
\usepackage[framemethod=TikZ]{mdframed}
\usepackage{xcolor}
\usepackage[shortcuts]{extdash} 

\usepackage{cite}

\IEEEoverridecommandlockouts
\begin{document}



\newenvironment{textbox}[2][]{%
\ifstrempty{#1}
{\mdfsetup{frametitle={}}}%
{\mdfsetup{innerleftmargin=10pt, innerrightmargin=10pt,%
frametitle={%
\tikz[baseline=(current bounding box.east), outer sep=0pt]
\node[anchor=east,rectangle,draw=black!75,fill=tablebg]
{\textcolor{black}{~#1~}};}}}%
\mdfsetup{innertopmargin=4pt,innerbottommargin=5pt, linecolor=black!75,%
linewidth=.5pt,topline=true,roundcorner=5pt,skipabove=5pt,%
frametitleaboveskip=\dimexpr-\ht\strutbox\relax}
\begin{mdframed}[]\relax%
\label{#2}}{\end{mdframed}}


\definecolor{pastelgreen}{HTML}{ccebc5}
\definecolor{pastelyellow}{HTML}{fed9a6}
\definecolor{pastelred}{HTML}{fbb4ae}
\definecolor{pastelrose}{HTML}{fddaec}
\definecolor{pastelgray}{HTML}{f2f2f2}
\definecolor{tablebg}{HTML}{f0f0f0}


\newcommand{\stackoverflow}{Stack Overflow}

\newcommand{\markmark}[1]{{\color{blue} Mark: #1}}
\newcommand{\andy}[1]{{\color{orange} Andy: #1}}
\newcommand{\td}[1]{{\color{red} #1}} 

\newcommand*\circled[1]{\tikz[baseline=(char.base)]{
\node[minimum size=\baselineskip,shape=circle,draw,inner sep=1pt,font=\footnotesize,fill=tablebg,] (char) {#1};}}

\newcommand\revised[2]{#1}

\title{Deconstructing Sentimental \stackoverflow{} Posts Through Interviews: Exploring the Case of \\ Software Testing*\thanks{* Accepted at the \href{https://conf.researchr.org/home/chase-2023}{\faIcon{link} CHASE 2023 Registered Reports Track}}}


\author{\IEEEauthorblockN{Mark Swillus}
\IEEEauthorblockA{\textit{Delft University of Technology} \\
The Netherlands \\
m.swillus@tudelft.nl}
\and
\IEEEauthorblockN{Andy Zaidman}
\IEEEauthorblockA{\textit{Delft University of Technology} \\
The Netherlands \\
a.e.zaidman@tudelft.nl}
}

%



\maketitle
\begin{abstract}
The analysis of sentimental posts about software testing on \stackoverflow{}
reveals that motivation and commitment of developers
to use software testing methods
is not only influenced by tools and technology.
Rather, attitudes are also influenced by socio-technical factors.
No prior studies have attempted to talk with \stackoverflow{} users about
the sentimental posts that they write, yet, this is crucial to understand
their experiences of which their post is only one fragment.
As such, this study explores the precursors
that make developers write sentimental posts
about software testing on \stackoverflow{}.
Through semi-structured interviews, we reconstruct the individual experiences of
\stackoverflow{} users leading to sentimental posts about testing.
We use the post as an anchor point
to explore
the events that lead to it and how users
moved on in the meantime.
Using strategies from socio-technical grounded theory (STGT),
we derive hypotheses about the socio-technical factors
that cause sentiment towards software testing.
\end{abstract}

\section{Introduction}
We already know for over 40 years that software testing
is one of the most pragmatic mechanisms
by which we can ensure
the quality of the software artefacts that
we engineer~\cite{carstensen_lets_1995,hetzel_complete_1988,myers_art_2012,yourdon_managing_1988}.
In the light of the unquestionable growing impact
that software and software supported devices are
having on our daily lives,
the role of software testing becomes ever more important.
However, to this day there is a schism
between widespread recommendations for software engineering practice
and our knowledge of how software testing \emph{actually} happens.
The urgency to solve this conflict
was also signalled by others
with a call to arms
to better understand the testing
process~\cite{bertolino_software_2007,mantyla_who_2012}.
We have recently seen studies emerge
that have observed
how software developers test.
Beller et al.~\cite{beller_developer_2019} have investigated
when and how
developers write test cases
in their Integrated Development Environment.
They observed
that around 50\% of the studied projects
do not employ automated testing methods at all.
But they also found out
that for almost all cases
testing happens far less frequently
than developers estimate.
If testing is truly considered a last line of defense
against software defects,
we need to understand why developers \emph{do} or \emph{do not}
engineer and execute test cases.
We have already seen glimpses of this in literature.
Studies have shown
that company culture
or time pressure
leads to cognitive biases during testing~\cite{calikli_influence_2013, mohanani_cognitive_2020, salman_what_2022},
estimations of the time it takes
to write test
are often inaccurate~\cite{beller_developer_2019, kasurinen_analysis_2009},
availability of documentation
shapes the development of tests~\cite{aniche_how_2022},
and that the cost/benefit of testing
is often unclear~\cite{begel_analyze_2014}.
Additionally,
Kasurinen et al.~\cite{kasurinen_analysis_2009}, Runeson~\cite{runeson_survey_2006}, and Daka and Fraser~\cite{daka_survey_2014}
highlight issues with motivating developers
to test software:
only half of them
have positive feelings about testing,
and approachability of tools is a
major factor.
Their research demonstrates that
technical and social aspects
which affect practitioners'
motivation and commitment to test
are interwoven.
We argue that one needs to consider
the interplay of technical and social aspects
to truly understand why practitioners do (not)
test.
As we set out to investigate what
influences software engineers
when practicing software testing,
we therefore go beyond an analysis of technical aspects
to reveal socio-technical factors
that influence practitioners in their decision-making.
\revised{%
In line with the definition of socio-technical systems
by Whitworth et al. \cite{Whitworth_social_2009},
we understand as socio-technical factors
the intertwined technical and social factors that
contextualize the creation of software artifacts.
\textit{``Social''} includes for example the
people, their interaction, company policies and norms.
\textit{``Technical''} in this context includes components of the technical infrastructure
that enable and support the creation of artifacts and the facilitation of social needs
(e.g., tools for communication).}{2A.4}

\revised{%
Understanding socio-technical factors
of software testing is important as
they directly influence the lived experience of practitioners
and have an influence on the quality of their produced artifacts.
Emotional attitudes and unhappiness for example,
which are very likely be connected to these factors
can have a detrimental effects
as they can for example lead to process divergence~\cite{graziotin_what_2018}.}{2A.1}
In the scope of an exploratory study we
have already identified factors
that are linked to attitudes about testing~\cite{swillus_sentiment_2023}.
In~\cite{swillus_sentiment_2023}, we have qualitatively analyzed 200 posts
and find that practitioners
who ask their questions on \stackoverflow{}
in a sentimental way,
show aspiration or report discouraging circumstances.
\revised{%
As \textit{sentimental} we consider expressions that indicate
emotional arousal of a person or statements that reflect opinionated views
and subjective, sometimes judgemental perspectives.}{2A.4}
Both positive and negative posts on \stackoverflow{}
show that sentimental questions are often asked
to discover new approaches or to reflect on practices.
Through our analysis we were able to pinpoint
reoccurring issues that are connected with negative and
positive sentiment.
For example,
like Pham et al. we find
that project complexity seems to be a factor
that influences decision making
and ignites attitudes about software testing~\cite{pham_enablers_2014}.
We investigated \textit{how} software engineers
express sentiment about testing.
With the knowledge we gained
we take the next step,
asking \textit{why} software engineers express sentiment
by illuminating the larger context
of posts that we analysed.
In this study
we explore this context
by interviewing authors,
reconstructing the individual experience
leading to sentimental posts.
\revised{%
We argue that deconstructing posts in this way
gives us insights into the lived experience of software developers
and their perception of software testing.
Analysing these experiences and perceptions helps us to understand
why developers decide (not) to test.
We approach the reconstruction of experiences by first
investigating how and why sentimental posts}{2B.1}
\==~in which authors ask their questions in a subjective
or emotional way~\==~
come about.
Together with the authors,
we aim to explore events and circumstances
that are related to their sentimental post.

\begin{textbox}[RQ1: Why are Practitioners Sentimental?]{}
    \begin{itemize}[leftmargin=28pt]
   \item[\textbf{RQ1.1}] Which events and circumstances lead software engineers to post about testing on \stackoverflow{}?
   \item[\textbf{RQ1.2}] What influences them to become sentimental?
   \end{itemize}
\end{textbox}

By analysing what practitioners tell us about
their experience,
we want to identify factors that influence
motivation, adoption and commitment to testing practices.

\begin{textbox}[RQ2: Motivation and Commitment to Test]{}
    Which factors affect software developers who ask sentimental questions on \stackoverflow{} (not) to use systematic software testing?
\end{textbox}

\revised{%
In our former work that investigated how practitioners
express themselves sentimentally on \stackoverflow{}~\cite{swillus_sentiment_2023}
we developed hypotheses that approach \textbf{RQ2}.
To guide our investigation into \textbf{RQ2}, we consider these hypotheses while remaining open
to other theoretical directions.
In \cref{sec:hypotheses} we present further details regarding the hypotheses.
A more detailed elaboration can be found in~\cite{swillus_sentiment_2023}.}{A2.5}
\newpage

\begin{samepage}
\begin{textbox}[Hypotheses to Guide Interviews for RQ2]{}
\begin{itemize}[leftmargin=15pt]
\item[\textbf{H$_{1}$}] The motivation to test depends on:
\begin{itemize}[leftmargin=12pt]
\item[\textbf{H$_{1.1}$}] The style of project management (socio-technical).
\item[\textbf{H$_{1.2}$}] The approachability of testing with best practices (technical).
\item[\textbf{H$_{1.3}$}] How peers communicate its use and value (social).
\end{itemize}

\item[\textbf{H$_{2}$}] If developers have never experienced the value of testing in larger, more complex projects, they will not be inclined to test.
\item[\textbf{H$_{3}$}] Adoption or learning of testing in complex projects provokes negative attitude towards it
\item[\textbf{H$_{4}$}] Adoption of or change in software testing is inspired and determined by shared, social experiences, and not so much by tools and technology.
\end{itemize}
\end{textbox}

\end{samepage}

Finally, we want to understand
how the attitude towards testing
and the way in which
it is carried out
changes over time
as participants do (not) gain
knowledge and confidence
about software testing
approaches through experience.
\revised{%
By analysing and comparing interviews
we want to identify the possible different
trajectories that software
developer can take as they change roles and responsibilities
regarding the development of software
and how it influences
they way in which they perceive testing.

\begin{textbox}[RQ3: Evolving Attitudes About Testing]{}
    How does the subjective perception and experience of software testing change over time?
\end{textbox}}{2B.3}

In semi-structured interviews we
use the \stackoverflow{} post of the interviewee as a chronologic anchor point.
We explore the events that lead to it,
and reflect together with the participants
how they moved on in the meantime.
By setting our analysis of posts into contrast
with the perspectives that participants explain to us in interviews
we derive hypotheses about socio-technical factors
that cause sentiment towards software testing.

\section{Hypotheses}
\label{sec:hypotheses}
%
Whether practitioners of software engineering
have a positive or negative attitude
towards testing depends on many individual factors.
Efficiency in testing, and the willingness to systematically test software
is not only a matter of tools and technology.
\revised{In our study of posts on \stackoverflow{}~\cite{swillus_sentiment_2023} we
find that a confrontation with challenging testing scenarios
can under some circumstances cause negative feelings.}{2A.5}
We know that developers distance themselves from tasks
to which their unhappiness relates~\cite{graziotin_what_2018}.
The confrontation with challenging scenarios
can therefore lead to withdrawal from testing, potentially
resulting in process deviation and reduced code quality.
However, we also see that
challenges can increase the
motivation or ambition of practitioners
in the case of software testing.
Creativity and being able to make a difference can make
the engineering of test suites
worthwhile, even when challenges arise.
Whether engineering of sophisticated test cases
leads to increased motivation or withdrawal
we therefore argue,
highly depends on context.

Additionally Meyer et al.~\cite{meyer_today_2021} found out that on good workdays,
developers make progress and create value for projects they consider meaningful.
On good days, they spend their time efficiently,
with little administrative work, and without facing infrastructure issues;
what makes a workday typical and therefore good is primarily assessed
by the match between developers’ expectations and reality.
We argue that hindrances to engineers who are working on test cases,
created for example by infrastructure issues,
overly complicated development environments,
or \textbf{constraints introduced through the project management style
affect practitioners' attitude towards testing~(H$_{1.1}$)}.
We hypothesise that \textbf{practitioners are motivated} to invest
their time into testing,
\textbf{when best-practices} or examples from documentations
\textbf{can easily be applied~(H$_{1.2}$)}.
Even more so if they and their \textbf{peers communicate the value that
systematic testing can bring to projects~(H$_{1.3}$)}.

Accordingly, practitioners who
already consider testing
a valuable building block of software engineering
are ambitious and aspirational about it.
We further hypothesize that
\textbf{the value of testing is not evident to practitioners
that never experienced its benefits in larger, complex projects~(H$_{2}$)}.
Observations of Pham et al.~\cite{pham_enablers_2014} seem to support this hypothesis.
They identified that novice developers
adjust their testing effort according to the perceived complexity of code.
A project has to be complex to warrant testing to be beneficial.
We suggest that there is a huge potential for
negative experiences here.
Practitioners only start to test when they perceive a project as complex enough,
but in those cases testing in not easily approachable anymore.
Complexity we argue
causes unexpected behaviors
and makes best-practices hard to apply.
If not supported by more experienced peers within a project,
\textbf{the adoption of testing in complex projects can provoke negative feelings
and potentially a consequential withdrawal from testing~(H$_3$).}
Pham et al.~\cite{pham_enablers_2014} even report that some students
develop an anxious attitude towards testing while they learn it.

Positive feelings or ambitions mentioned in posts on \stackoverflow{}
are often self-aroused for example through engagement with
\textit{inspiring} resources like books or blogs.
Daka and Fraser also identified that
peer pressure is only rarely mentioned
as a motivating factor to write unit tests;
the driving force for a developer to use unit testing
is supposedly 
their own conviction~\cite{daka_survey_2014}.
We contend however that project specific non-technical factors
like the knowledge of testing within a team
and the way in which developers contribute to the project
do play an important role.
Changing of attitude about testing,
so we hypothesize,
may be stimulated by the adoption of new testing tools,
but more crucially,
\textbf{change is inspired by human and social interaction
and determined by \textit{shared} experiences~(H$_4$)}.

\section{Participants and Dataset}
\label{sec:dataset}

In a prior study we qualitatively
analyzed 200 \stackoverflow{} posts about software testing~\cite{swillus_sentiment_2023}.
\stackoverflow{} posts are living documents
which are edited by their authors and moderators
and extended with comments
sometimes even a decade after they were created.
We analyzed if and how practitioners express sentiment
about software testing in posts
and what the circumstances are that
practitioners describe when they express sentiment.
We considered posts to be sentimental
when questions are asked in an emotional,
subjective way,
demonstrating that their author
is in some way aroused,
for example when authors indicate
an aversion or attraction
to testing.
The replication package which contains all posts
and the output of our work
is publicly accessible~\cite{swillus_replication_2022}.
Of the 200 posts that we manually analysed,
108 turned out to be sentimental,
reflecting negative, positive or both attitudes.
Based on our analysis
we constructed 22 focused
codes and four analytical categories.
Subsequently, we assigned codes and categories to all posts.
This has enabled us to
work with the dataset systematically
by comparing data and establishing connections between
underlying themes in the dataset. 
Coding and categorization illuminate
what practitioners write about when they are sentimental,
how they express their sentiment.
Our exploratory work also already
explores the reasons for sentimentality in posts.

For our interviews we recruit subjects from the list of authors
who wrote the posts we analyzed.
During interviews posts serve as chronological anchor points
to explore the events leading to them
and how attitudes, experience and other factors
have changed in the meantime.
Beyond serving as a chronological anchor point,
posts provide us thematic entry points
to explore details of individual circumstances.
By deepening our understanding of practitioners` lived experiences
through interviews
we will refine the preliminary findings of our former work.
We for example argue that
the practitioner writing the following post
has an aspirational attitude but expresses both a positive and negative
sentiment about testing.
\textit{“I’m refactoring one big complicated piece of code [...].
So, I need to write a unit test [...]. After googling I came up with 2 ideas [...].
Am I missing some silver bullet? Possibly, DBUnit is the tool for this?”}
We hypothesise that even when aspirational,
practitioners can develop negative attitudes because they face
their ambiguities about testing too late.
They get motivated to use testing only when the complexity of their project has
reached a threshold that is very hard to overcome without experience in testing.
In an interview we could for example
explore the post mentioned above further.
Focusing the interview on the circumstances that
required refactoring of the code
and why the necessity of testing arose in this context
will allow us to put our analysis into a bigger context.
In addition to the refinement of our preliminary work,
we thereby explore new vantage points for theory construction
to answer \textbf{RQ1\==3}.

As we approach data collection and analysis within the framework
of socio-technical grounded theory (STGT)~\cite{hoda_socio-technical_2022},
incorporating strategies from constructivist grounded theory,
we follow Charmaz' recommendations
not to pre-determine the sample size that is required
for us to reach theoretical saturation~\cite{charmaz_constructing_2014}.
Instead, using theoretic sampling
we alternate between data collection and data analysis
to deepen, refine and test the construction of
analytical categories, codes or interpretive theory.
We will therefore continue to conduct interviews
until our analysis reaches a point at which additional data
does not provide new insights in the form of codes or categories anymore.
As we approach data gathering and analysis in this iterative way,
we continuously recruit new subjects
and contact subjects again for follow up interviews
if we realize that more details
about concrete aspects are needed.

Summarized, the dataset that we will construct
and make available in form of a replication package
will contain the following artifacts:

\begin{textbox}[Artifacts to be Published]{}
\begin{itemize}[leftmargin=0pt]
    \item [] Anonymized interview transcripts
    \begin{itemize}[leftmargin=12pt]
        \item Initial interviews
        \item Follow-up interviews
    \end{itemize}
\item [] REFI-QDA\footnote{REFI-QDA is an open standard that
    enables interoperability
    between qualitative data analysis software: \href{https://www.qdasoftware.org/}{\faIcon{link} QDASoftware.org}}. file containing analysed dataset
    \begin{itemize}[leftmargin=12pt]
        \item Coded transcripts
        \item Categorization
        \item Contextual information (analytical memos)
    \end{itemize}
    \item [] Codebook
    \begin{itemize}[leftmargin=12pt]
        \item Description of codes
        \item Inclusion and exclusion criteria
        \item Examples for application of codes
    \end{itemize}
\end{itemize}
\end{textbox}

\section{Execution Plan}
\begin{figure*}[] 
	\centering
        \includegraphics[width=.8\textwidth]{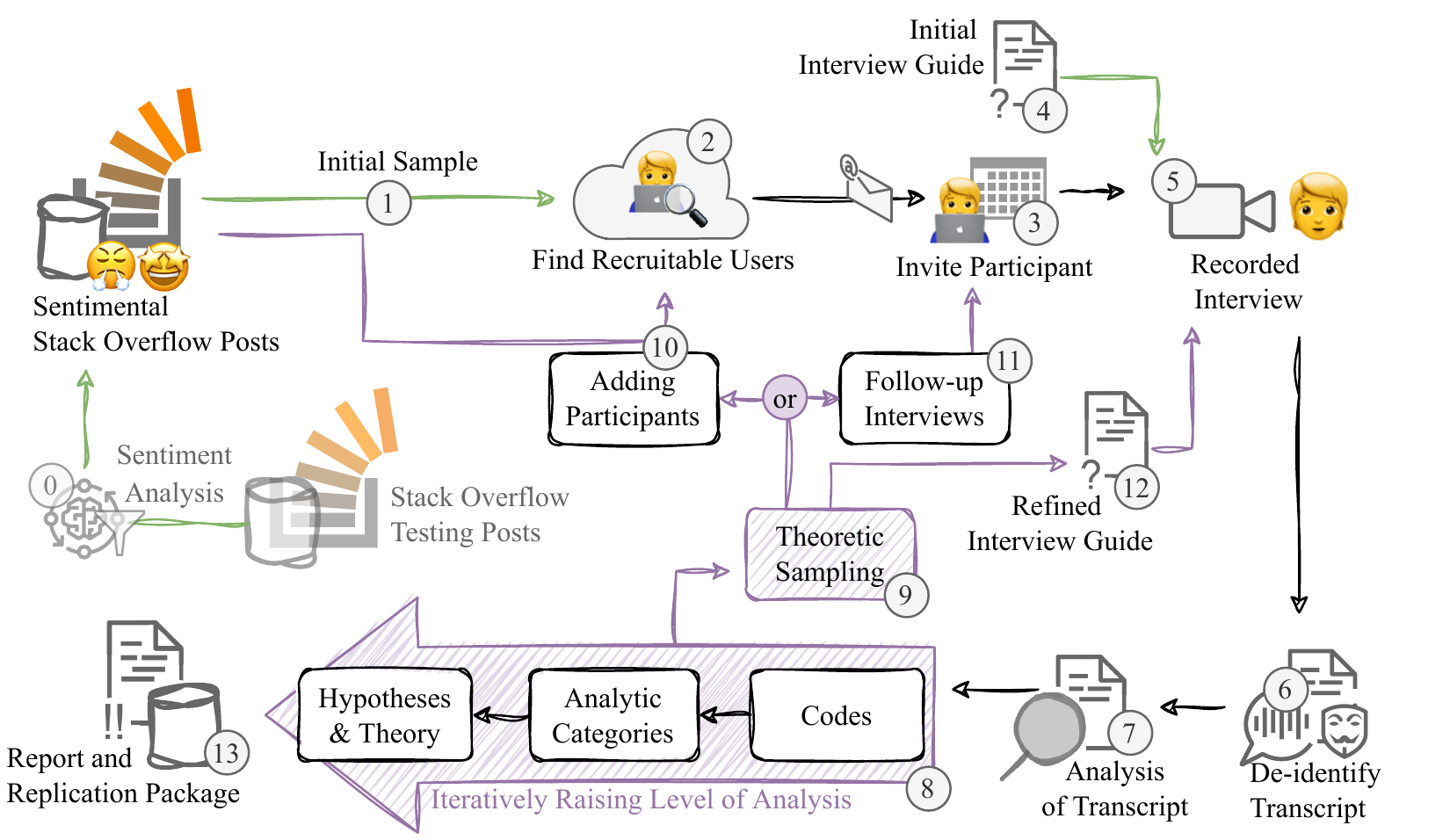}{}
        \caption{Execution plan that starts with recruitment
            of an initial group of participants \protect\circled{3}
            who are select
            from a dataset of sentimental Stack Overflow posts
            \protect\circled{1},
            and ends with
            the publication of a report \protect\circled{13}
            that contains new hypotheses and a theory
            we construct through systematic, iterative
            data analysis \protect\circled{8}.
            We consider additional samples from the \stackoverflow{}
            data dump \protect\circled{0} to protect participants' integrity
            and to support our recruiting approach in case of low turnout.
        }\label{fig:execution_plan}
\end{figure*}
\label{sec:execution}
At the core of our study we carry out and analyse
semi-structured interviews with software developers
that we recruit from sentimental \stackoverflow{} posts.
The analysis of these interviews gives
us insights into the lived experience
of software developers with a particular focus on how
they perceive and experience software testing.
Steps \circled{1} to \circled{12} in \Cref{fig:execution_plan}
illustrate how we prepare, conduct and analyse interviews
to systematically infer knowledge from
insights that developers give us.
The circular pattern in \Cref{fig:execution_plan}
which starts with \textit{theoretic sampling} \circled{9}
reflects the research methodology of
Socio-technical Grounded Theory (STGT)~\cite{hoda_socio-technical_2022}
which we use in our study.
Iteratively it leads us back to the analysis
of additional, focused interviews \circled{7}
and their incorporation into emerging codes,
categories and hypothesis until we hopefully reach a point
of saturation at which we report our findings and possibly
an interpretive theory that explains observed phenomena.

\subsection{Recruiting}%
\label{sub:de-anonymisation}
All contributions to \stackoverflow{} are openly accessible under the
creative commons license,
which makes \stackoverflow{} very accessible for data-mining.
The platform does however not provide a way to
contact users directly.
We therefore first need to identify users
which we are able to recruit via email \circled{2}.
\revised{%
Considering users' right to privacy and self-determination,
we only recruit users who publish their contact information
in a way that implies that they are expecting that this information
can be used by researchers.
For example, many users reference their personal website
which often contains a \textit{Contact} page
that indicates how the user can be reached
in a way that is not unsolicited.
If users do not provide any such references in their \stackoverflow{} profile,
we refrain from using extant data to de-anonymise users
as we and others~\cite{gold_ethics_2021}
consider practices to be unethical,
as they threaten the contextual integrity of participants~\cite{nissenbaum_contextual_2011}.}{2C.4}

\revised{%
    Collecting a substantial amount of data
    compensates the negative effects
    that misleading or fabricated accounts of participants cause.
    However, going into details too far
    by gathering and analysing too much data
    can lead to descriptivism where abstraction of
    observed phenomena is no longer possible~\cite{charmaz_constructing_2014}[p. 89].
    Accurately pre-determining the number of interviews that we need
    to conduct is therefore difficult.}{2A.5}
For our first round of interviews,
we aim to recruit 15 participants,
selecting 5 negative posts, 5 positive posts, and 5 post with both
negative and positive sentiment about testing.
\revised{%
    During this first round of interviews we will evaluate
    if those numbers are appropriate.}{2A.5}

\revised{%
In order to find recruitable participants we first consider the 200 posts
that we have analysed in our former study \circled{1}.
In case the dataset of 200 posts does not provide us enough potential candidates
for recruitment or because of ethical considerations on which we elaborate in \Cref{sec:ethics},
we will consider additional testing related posts from
the whole stack overflow data dump,
categorizing them by sentiment using sentiment analysis tools \circled{0}.}{2A.1}
Once we find participants that confirm their participation,
we invite them to select a date for their interview
and provide them additional information about our intentions
and the conditions of their participation to ask them for
their informed consent~\circled{3}.

\subsection{Semi-Structured Interviews}%
\label{sub:semi-structured interviews}
We are still in the process of exploring
the lived experiences and attitudes of practitioners
which includes their use of language.
We still need to learn how they refer to
and conceptualize \textit{software testing}
which is why we do not exactly know how participants
will interpret interview questions (yet).
To avoid imposing our own ideas and
our language onto participants,
especially during the first rounds of interviews,
we therefore avoid asking questions that are too leading.
This is especially important to avoid a confirmation bias
for testing of \textbf{H$_1$-H$_4$}.
Quite the opposite to imposing our theories,
we need to be able to follow unanticipated specifics,
hints, and views to gather detailed and genuine accounts of participants'.
This is fundamentally different from structured interviews
in which the researcher asks the exact same question to all participants.
To plan and guide our interviews,
we therefore base our strategy for semi-structured interviews
on Charmaz` \textit{intensive interviewing} method
which is designed as a tool to explore a person's substantial
experience with a research topic.
\revised{%
By following Charmaz' guidelines for intensive interviewing we also
guard ourselves against confirmation bias regarding the investigation of \textbf{H$_1$-H$_4$}.
In intensive interviewing the interviewer
creates room for the interviewee
to tell their story.
Instead of dictating the direction of the interview
through strictly following a detailed interview guide,
the interviewer uses soft control to fill out the details
of the story shared by the interviewer~\cite[p. 69]{charmaz_constructing_2014}.}{2A.3}

According to the method,
the initial interview-guide
that we create for our first round of interviews
will only be used to apply a soft
control in interviews which allows us to gently
keep participants on topic~\circled{4}.
As our study progresses and our
understanding of practitioners' experiences
and their use of language
grows, we refine the interview-guide,
adding pertinent questions that allow us to
systematically raise the level of our analysis~\circled{12}
to answer \textbf{RQ1-RQ3}.
All interviews are recorded so that the
interviewer can focus their attention on the subject,
without needing to jot down notes and to avoid mistakes
when reconstructing the interviews`
content from memory~\circled{5}.
After each interview the recording is transcribed
and anonymised, after which the recordings are destroyed to ensure
that the privacy of participants is protected~\circled{6}.
Once the transcription process is done we will
analyse the transcripts~\circled{7}.

\subsection{Systematic Data Analysis}%
\label{sub:systematic data analysis}
Through our qualitative analysis of posts
on \stackoverflow{} we learned that practitioners`
sentiments towards software testing are not caused
merely by technical aspects~\cite{swillus_sentiment_2023}.
The processes around the topic of software testing
is socio-technical in the sense that technical aspects
of practice are interwoven
with social aspects.
One needs to take both into consideration
to understand
why practitioners become sentimental about testing
and why they decide (not) to test the software they develop.
To lead our investigation of socio-technical aspects,
we employ strategies from Hoda's
Socio-Technical Grounded Theory (STGT)~\cite{hoda_socio-technical_2022}.
Using codes, categories and the preliminary interpretive theory,
that we constructed
in our prior study~\cite{swillus_sentiment_2023},
using strategies from STGT's \textit{Basic Stage}
for data collection and analysis,
we now focus on what Hoda defines as STGT's \textit{Advanced Stage}.
Concretely we use the \textit{emergent mode} for data analysis
and theory construction.
Working in the emergent mode for STGT studies
enables the emergence of theory
through iterative data collection
and analysis.
The emergent mode leads to theoretical saturation
and results in a mature theory
that is grounded
in the data that was collected~\cite[p. 14]{hoda_socio-technical_2022}.
Our epistemological stance with regard to our
research questions is that the
lived experience of testing practices and the experience of
practitioners is highly individual and constructed within a unique
complex socio-economic and technical context.
We cannot expect that participants fully comprehend all factors
that create and form their experience, even less
can we expect them to reconstruct all those factors in interviews with us.
What they share with us in interviews is a subjective
report of what they believe their experience constitutes.
With a pragmatist attitude,
we do think however that
our systematic approach will enable us to trace
factors that affect the lived experience of software engineers
when they use software testing.
Within the framework of Hoda's STGT we therefore
adopt a constructivist epistemology,
acknowledging that our work can only be an interpretive
translation of the complex lived experience
that practitioners describe to us.
To reduce the effect that our own ideas have on the study results,
we conduct interviews and data analysis with
the necessary scientific rigour,
by following systematic methods
from Charmaz' version of constructivist Grounded Theory~\cite{charmaz_constructing_2014}.

We begin the analysis
of transcripts by coding them line by line
using \textit{open coding}~\circled{8}.
After the initial round of about 15 interviews is finished,
we proceed with additional rounds of coding,
summarizing and refactoring codes,
also incorporating codes of our former study~\cite{swillus_sentiment_2023}.
Emerging themes, categories and \textit{focused codes} at this point
will then guide us in the process of theoretic sampling~\circled{9}.
\revised{%
To reduce the likelihood of a misinterpretation
that would pose a threat to the validity of
our results, the authors will discuss the interpretation of the data
recorded in memos and developed codes, categories and theory
and resolve disagreements in a cooperative manner.
We will not provide a quantitative analysis
of this process of reliability verification
as such an analysis would suggest a level
of objectivity that we do not want to claim~\cite{mcdonald_reliability_2019}.}{2A.3}
To deepen our understanding of emerging ideas,
we conduct further, more focused follow-up interviews
with participants~\circled{11} or
with new participants~\circled{10}.
\revised{%
    Follow-up interviews will enable us to go more into detail with participants
    while the recruitment of new participants enables us to broaden theory development.}{2A.5}
We create new interview guides for these interviews
to be able to focus on specifics
that are relevant to further develop our analysis
in order to answer \textbf{RQ1-RQ3}.
In this way, the iterative process of theoretic sampling
enables us to systematically test and extend our hypotheses.
To make this process of data analysis and theory construction systematic,
we employ methods for \textit{constant comparison} from
Charmaz\cite{charmaz_constructing_2014} and Saldaña~\cite{saldana_coding_2013}.
We compare for example statements and events within one interview,
statements from the same interviewee in different interviews,
or statements of different interviewees about similar incidents.
As we raise the level of analysis,
we also compare, on a more abstract level focused codes
or analytical categories that were assigned to interview segments.
We also incorporate the content of analytical memos
that we write during the whole process
of data analysis.
To facilitate the process of constant comparison for theory construction we use
techniques for qualitative research
like diagramming~\cite{charmaz_constructing_2014}
and clustering~\cite{saldana_coding_2013}.

\subsection{Reporting of Results}%
\label{sub:Reporting of Results}

The execution plan of our study
centers around the iterative process
of theoretic sampling and methods for constant
comparison which, using the framework of STGT
lead to theory development.

We start with a sentimental post
of a \stackoverflow{} user, collect data about
the context of that post through interviews
and analyse this data.
As we increase the level of analysis,
we supplement the dataset of anonymised transcripts
by assigning codes and categorization
and by writing analytical memos.
In our report we lead the reader through this process
to make transparent how we moved from
a low level of analysis at which we construct codes
that describe and summarize segments
of interviews to a higher level of analysis,
that concludes with the construction of a theory.
For each level of analysis our report provides concrete
examples of application,
demonstrating for example
how codes were applied to interview segments and
how the categories emerged.
Apart from making our approach transparent in the report,
we present answers to \textbf{RQ1-RQ3},
and show how the research questions
and the hypotheses \textbf{H$_1$-H$_4$}
guided our analysis.
We also elaborate on the testing of
hypotheses \textbf{H$_1$-H$_4$}
and suggest arguments for their rejection, validation or refinement.

Finally, the reporting of the theory
which we developed
is central to the publication of the results
of this study.
There is however a difficulty here:
it is impossible to foresee the depth
and maturity of such a theory,
before we start our analysis.
We might for example realize that
the research population or the data gathering
method used in this study is not sufficient
to reach theoretic saturation that is required
to report a mature theory.
A conclusion could be that
more triangulation of method or data
is needed in order to establish maturity~\cite{storey_who_2020}.
However, as Hoda emphasizes,
reporting of preliminary theories and
emerging hypotheses~\==~like we did in our
former study~\cite{swillus_sentiment_2023}~\==~is important to
assess and improve the relevance and rigour~\cite{hoda_socio-technical_2022}.
The reporting of our results will thus,
even in the case that it
centers around the presentation
of a \textit{preliminary theory}
motivate new venture
points to investigate socio-technical
aspects of software engineering which
will help us to better understand and
ultimately enrich the experience of software developers.

\medskip

To stimulate future research
and to subscribe to the values of open science,
we publish all anonymised data and
our research
output~\==~as outlined in \Cref{sec:dataset}~\==~openly
in the form of a replication package.

\revised{%
\section{Ethical considerations}\label{sec:ethics}
This study investigates and reports an analysis of perspectives of human subjects.
We want to ensure that participants are not harmed through our study.
Neither directly through the recruitment- or data collection process,
nor indirectly through repercussions caused by the publication of our work and its artifacts.
We also want to respect the policies of online platforms
when we extract data to identify subjects for recruitment.
For example, we are carefully considering case by case whether to use email addresses
published by users on GitHub profile pages
as it is stated in GitHub's
accepted-use policy that extracting email addresses to send unsolicited emails
is not permitted\cite{gitHub_github_2023}.

To protect subjects from harm
we consider their right to privacy and self-determination
and follow Nissenbaum's principles to protect contextual integrity~\cite{nissenbaum_contextual_2011}.
For recruitment, we only consider information that has been
made public by subjects in a way that implies an expectation
that the information can be used by researchers to contact them.
This includes for example email addresses referenced on the \textit{Contact}
page of a personal website
which has been linked by the user on their public \stackoverflow{} profile.
We also consider Marwicks’s concerns about context collapse\cite{marwick_i_2011}.
Users on social media websites like Stack Overflow participate and contribute
within a specific social context that is largely defined
by their expectations of how their contributions are perceived and used.
Linking these contexts,
for example by establishing a link between activities on Stack Overflow
and profiles on Instagram or Twitter through de-anonymization
leads to a collapse of this context.
Connecting content that individuals contributed in different contexts
can cause unexpected and damaging consequences.
We therefore de-identify all information that participants provide us before publication.
To reduce the lieklihood of the reversal of this de-identification,
we won't make transparent
which samples of the dataset were considered for recruitment of participants.
Additionally, after reviewing the dataset from which we select potential participants,
we will evaluate how likely a reversal of de-identification is
considering the group- and dataset size.
In case we cannot be confident in the de-identification approach
at this point,
we will take additional measures
and adjust our recruiting strategy
to reach that confidence.

To seek balance between transparency and protection of subjects’ integrity when
publishing our results,
we will ask participants to review the de-identified transcripts
and allow them to remove
the parts that they are not comfortable to share before
publication.

\medskip

Our study design was submitted and approved by the privacy team and ethics council of TU Delft.
}{2C.4}

\section*{Acknowledgements} 
This research was partially funded by the Dutch science foundation NWO through the Vici ``TestShift'' grant (No. VI.C.182.032).

\balance
\bibliographystyle{IEEEtran}
\bibliography{references}

\end{document}